\documentclass[amsmath,amssymb,showkeys,superscriptaddress,aps,prd,10pt,onecolumn]{revtex4-2}
\usepackage{graphicx}
\usepackage[utf8]{inputenc}
\usepackage[caption=false]{subfig}
\usepackage{multirow}
\usepackage{appendix}
\usepackage{graphicx,epstopdf}

\usepackage{colortbl}
\usepackage{xcolor}
\usepackage[colorlinks=true, urlcolor=blue, linkcolor=blue, citecolor=blue]{hyperref}
\usepackage{float}

\begin{document}

\title{Magnetic Moment of $\Xi_b(6227)$ as Molecular Pentaquark State}
\author{Halil Mutuk}%
\email[]{hmutuk@omu.edu.tr}
\affiliation{Department of Physics, Faculty of Sciences, Ondokuz Mayis University, 55200, Samsun, Türkiye}

 
\begin{abstract}
Motivated by the observation of $\Xi_b(6227)$ state, in this study considering $\Xi_b(6227)$ has a molecular structure, we calculate magnetic moment of this state in quark model. The magnetic moment of a hadron gives valuable information about the internal structure and shape deformations. We observe that orbital excitation of $\Xi_b(6227)$ molecular state change the results of magnetic moment significantly. We also observe that light quarks in $\Xi_b(6227)$ molecular state determine magnetic moment. Measurement of the magnetic moment of $\Xi_b(6227)$ can clarify the nature of this state and  be useful to identify the quantum numbers.
\end{abstract}

\maketitle

\section{Introduction}\label{introduction}
Singly heavy baryons have an important role in the hadron spectrum since the mass difference between the heavy quark (charm or bottom) and light quarks is large which makes heavy quark symmetry palpable for these systems. The presence of one heavy quark presents a laboratory for studying light quark dynamics in terms of heavy quark symmetry. Furthermore, singly heavy baryons may help to clarify the nonperturbative nature of the QCD. 

In 2018, the LHCb Collaboration reported observation of $\Xi_b(6227)^-$ baryon in both the $\Lambda^0_b K^-$ and $\Xi_b^0 \pi^-$ decay modes with  mass $m_{\Xi_b }= 6226.9 \pm 2.0 (\mbox{stat}) \pm 0.3 (\mbox{syst}) \pm 0.2~\mbox{MeV}$ and width $\Gamma_{\Xi_b }= 18.1 \pm 5.4 (\mbox{stat}) \pm 1.8 (\mbox{syst})$~\cite{LHCb:2018vuc}. The observed mass and decay modes suggest that $\Xi_b(6227)^-$ could be a 1P or 2S excited state. After the observation of this state, many studies are done to identify the characteristics of this baryon. Refs. \cite{Chen:2018orb,Chen:2018vuc,Wang:2018fjm,Aliev:2018lcs,Nieves:2019jhp,Cui:2019dzj,Jia:2019bkr,Azizi:2020azq,Jia:2020vek,Yang:2020zrh,Yu:2021zvl,He:2021xrh,He:2021xrh,Kakadiya:2022zvy,Wang:2022dmw} considered this state as the excited of $\Xi_b$ baryon, whereas Refs. \cite{Huang:2018bed,Yu:2018yxl,Wang:2020vwl,Zhu:2020lza,Ozdem:2021vry} considered as a pentaquark state. Despite the conducted experimental and theoretical studies, the spin-parity of this state is still unknown. Furthermore even though the theoretical studies which are done as excited conventional baryon for $\Xi_b(6227)^-$, (for example $1P$-wave $\Xi^{'}_b$ with $J^{P}=3/2^{-}$~\cite{Kakadiya:2022zvy,Chen:2018orb,He:2021xrh,Cui:2019dzj,Wang:2018fjm,Yang:2020zrh}, $1P$-wave $\Xi^{'}_{b}$ with $J^{P}=5/2^{-}$~\cite{Chen:2018orb,Wang:2018fjm}, $1P$-wave or $2S$-wave with  $J^{P}=3/2$~\cite{Azizi:2020azq}), $\Xi_b(6227)^-$ might be a $\bar{K}\Sigma_b$ hadronic molecular state, due to the fact that the mass gap between the excited state $\Xi^*_b$ baryon and the ground state $\Xi_b$ baryon is about 440 MeV which is large enough to excite a light quark-antiquark pair to form a molecular state. In accordance with this argument, in Ref. \cite{Huang:2018bed}, strong decays of $\Xi_b(6227)^-$ as a $\Sigma_b \bar{K}$ molecule are studied with possible spin-parity assignments $J^P=1/2^{\pm}$ and $J^P=3/2^{\pm}$. They found that $J^P=1/2^-$ spin-parity assignment is favorable according to LHCb data. Under assumption of meson-baryon interaction, local hidden gauge approach in the Bethe-Salpeter equation was used to study mass of some $\Xi_c$ and $\Xi_b$ states \cite{Yu:2018yxl}. They obtained two poles with masses and widths significantly close to the $\Xi_b$ experimental data for both $J^P=1/2^-$ and $J^P=3/2^-$ quantum numbers. In Ref. \cite{Wang:2020vwl}, masses and pole residues of $\Xi_b$ as molecular pentaquark states with $J^P=1/2^{\pm}$ quantum numbers are obtained using QCD sum rule approach. Their results support $J^P=1/2^{-}$ molecular pentaquark picture for $\Xi_b(6227)$. Radiative decays of $\Xi_b(6227)$ as a molecular state of $\Xi_b$ baryon and $\bar{K}$ meson with $J^P=1/2^{-}$ quantum number are studied \cite{Zhu:2020lza}. They concluded that an experimental determination of the radiative decay width for $\Xi_b(6227)$ is needed to understand inner structure. In Ref. \cite{Ozdem:2021vry}, the magnetic moment of $\Xi_b(6227)$ is obtained in molecular scenario by using light-cone QCD sum rule.

As can be seen in studies cited above, there is no consensus on the internal structure of this particle. In addition to spectroscopic properties such as mass, decay widths and decay channels, other physical properties should be studied to elucidate internal the structure of $\Xi_b(6227)$. In this respect, investigation of magnetic moment of this particle may serve to enlighten inner structure of $\Xi_b(6227)$. Electromagnetic properties of hadrons have an important role on strong interactions and provide insight into low-energy regime of QCD. In accordance, magnetic moment contains important information about inner structures and shape deformations of the related hadron. In addition to this, magnetic moment may present valuable information about charge distribution and magnetization. 

Motivated by this situation, we study magnetic moment of $\Xi_b(6227)$ assuming molecular pentaquark picture in quark model. The paper is outlined as follows: in Section \ref{formalism} we obtain wave functions of $\Xi_b(6227)$ in quark model considering it as a molecular state. The numerical analysis together with the discussion is given in Section \ref{numerical}. Section \ref{final} is reserved for the summary.

\section{$\Xi_b(6227)$ Wave function } \label{formalism}
There are four degrees of freedom in the quark level for a given particle: color, spin, flavor and space. Using these degrees of freedom, the total wave function for a multiquark state can be written as
\begin{equation}
\Psi_{total}=\phi_{flavor}\chi_{spin}\xi_{color}\psi_{space},
\end{equation}
where the total wave function should be antisymmetric due to the Fermi statistics. We study $\Xi_b(6227)$ state in a $SU(3)_f$ frame and accordingly construct the flavor wave function. A pentaquark state in molecular picture is made up of a baryon and a meson. The quark decomposition of $\Xi_b(6227)$ in molecular picture is assumed to be as $(\bar{u}s)(udb)$. The two light quarks $(ud)$ in the singly bottomed baryon can be symmetrical or antisymmetrical. In the symmetrical configuration, singly bottomed baryon belongs to $6_f$ flavor representation and it can form $10_f$ and $8_{1f}$ with the stranged meson $(3_f)$. In the antisymmetrical configuration, singly bottomed baryon belongs to $\bar{3}_f$. In this case, it can form $8_{2f}$ and $1_f$ with stranged antimeson $(3_f)$. As a result, the flavor representation turns out to be $3 \otimes  3 \otimes 3= 1 \oplus 8_1 \oplus 8_2 \oplus 10$. Since we assume that $\Xi_b(6227)$ is a molecular state of a baryon and a meson, corresponding flavor and spin wave functions of baryon and meson should be given. In Tables \ref{tab:baryon} and \ref{tab:meson}, we present wave functions of $\Sigma_b$ baryon and $K$ meson, respectively.

\begin{table}[h]
\caption{\label{tab:baryon}The flavor $\phi_{flavor}$ and spin $\chi_{spin}$ wave functions of $\Sigma_b$ baryon in terms of $S$ and $S_3$ where $S$ represents spin and $S_3$ represents its third component.}
\begin{ruledtabular}
\begin{tabular}{lcr}
Baryon &$(S,S_3)$&Flavor $\otimes$ spin\\
\hline\\
$\Sigma_b^{-}$& $\vert{\frac{1}{2},\frac{1}{2}} \rangle$   &${\frac{1}{\sqrt{2}}}(dsb - sdb) \otimes {\frac{1}{\sqrt{2}}}(\uparrow\downarrow\uparrow-\downarrow\uparrow\uparrow)$ \\ \\

& $\vert{\frac{1}{2},-\frac{1}{2}} \rangle$   &$\frac{1}{\sqrt{2}}(dsb - sdb) \otimes {\frac{1}{\sqrt{2}}}(\uparrow \downarrow \downarrow -\downarrow\uparrow\downarrow)$
 \\ \hline \\
 
$\Sigma_b^{0}$& $\vert{\frac{1}{2},\frac{1}{2}} \rangle$   &${\frac{1}{\sqrt{2}}}(usb - sub) \otimes {\frac{1}{\sqrt{2}}}(\uparrow\downarrow\uparrow-\downarrow\uparrow\uparrow)$ \\ \\

& $\vert{\frac{1}{2},-\frac{1}{2}} \rangle$   & ${\frac{1}{\sqrt{2}}}(usb - sub) \otimes {\frac{1}{\sqrt{2}}}(\uparrow\downarrow\downarrow-\downarrow\uparrow\downarrow)$  \\ \hline \\

$\Lambda_b^{0}$& $\vert{\frac{1}{2},\frac{1}{2}} \rangle$   &${\frac{1}{\sqrt{2}}}(udb - dub) \otimes {\frac{1}{\sqrt{2}}}(\uparrow\downarrow\uparrow-\downarrow\uparrow\uparrow)$ \\ \\

& $\vert{\frac{1}{2},-\frac{1}{2}} \rangle$   &${\frac{1}{\sqrt{2}}}(udb - dub) \otimes {\frac{1}{\sqrt{2}}}(\uparrow\downarrow\downarrow-\downarrow\uparrow\downarrow)$

\end{tabular}
\end{ruledtabular}
\end{table}

\begin{table}[h]
\caption{\label{tab:meson}The flavor $\phi_{flavor}$ and spin $\chi_{spin}$ wave functions of $K$ meson in terms of $S$ and $S_3$ where $S$ represents spin and $S_3$ represents its third component.}
\begin{ruledtabular}
\begin{tabular}{lcr}
Baryon &$(S,S_3)$&Flavor $\otimes$ spin\\
\hline\\
$\bar{K}^0$& $\vert 0,0 \rangle$   &$\bar{d} s \otimes {\frac{1}{\sqrt{2}}}(\uparrow\downarrow-\downarrow\uparrow)$ \\ \hline \\

$K^-$&  $\vert 0,0 \rangle$  &$\bar{u} s \otimes {\frac{1}{\sqrt{2}}}(\uparrow\downarrow-\downarrow\uparrow)$
 \\ \hline \\
 
& $\vert 1,1 \rangle$   &$\bar{d} s \otimes \uparrow\uparrow$ \\ \\ 

$\bar{K}^{(*)0}$ & $\vert 1,0 \rangle$   &$\bar{d} s \otimes {\frac{1}{\sqrt{2}}}(\uparrow\downarrow + \downarrow\uparrow)$ \\ \\ 

& $\vert 1,-1 \rangle$   &$\bar{d} s \otimes \downarrow\downarrow$ \\ \\ \hline

& $\vert 1,1 \rangle$   &$\bar{u} s \otimes \uparrow\uparrow$ \\ \\ 

$\bar{K}^{(*)-}$ & $\vert 1,0 \rangle$   &$\bar{u} s \otimes {\frac{1}{\sqrt{2}}}(\uparrow\downarrow + \downarrow\uparrow)$ \\ \\ 

& $\vert 1,-1 \rangle$   &$\bar{u} s \otimes \downarrow\downarrow$ \\ \\

\end{tabular}
\end{ruledtabular}
\end{table}

Using Clebsch-Gordon coefficients, we obtain possible flavor wave functions of $\Xi_b(6227)$ as a molecular pentaquark state which can be seen in Table \ref{tab:flavor}.

\begin{table}[h]
\caption{\label{tab:flavor}The flavor wave functions of $\Xi_b(6227)$ as a molecular pentaquark state.  $I$ represents isospin and $I_3$ represents its  third component.}
\begin{ruledtabular}
\begin{tabular}{lcr}
Multiplet &$(I,I_3)$&Flavor wave function\\
\hline\\
$8_{1f}$& $(\frac{1}{2},-\frac{1}{2})$ & $  \sqrt{\frac{1}{3}}\left|\Sigma_b^{-} \bar{K}^{(*)0}\right\rangle-\sqrt{\frac{2}{3}}\left|\Sigma_b^{0}K^{(*)-}\right\rangle$  \\ \hline \\
$8_{2f}$& $(\frac{1}{2},-\frac{1}{2})$ & $ \left|\Lambda_b^{0} K^{(*)-}\right\rangle$ \\
\end{tabular}
\end{ruledtabular}
\end{table}

All the hadrons, whether conventional or exotic, should be colorless. As mentioned before, a molecular pentaquark state is composed of meson and baryon as subconstituents. In the molecular pentaquark state, the meson and baryon are in the color singlet. The color representation of meson in pentaquark tends to be $\bar{3}_{c} \otimes3_{c}=1_{c}\oplus8 _{c}$, whereas the color representation of baryon in pentaquark tends to be $3_{c} \otimes(3_{c}\otimes3_{c})=3_{c} \otimes(\bar3_{c}\oplus6_{c})=(1_{c}\oplus8_{2c})\oplus(8_{1c}\oplus10_{c}) $.

The quantum numbers of $\Xi_b(6227)$ as a molecular pentaquark  can be constructed by the coupling of baryon spin and meson spin. Denoting  $ J^{P} $ as the total spin of pentaquark state, we can write $ J^{P} = J_{b} ^{P_{b}}\otimes J_{m} ^{P_{m}}$ where $J_{b} ^{P_{b}}$ and $J_{m} ^{P_{m}} $ are corresponding to the angular momentum and parity of baryon and meson, respectively. There are three spin configurations we study: $J^P=\frac{1}{2}^{-}(\frac{1}{2}^{+}\otimes0^{-})  $, $J^P=\frac{1}{2}^{-}(\frac{1}{2}^{+}\otimes1^{-})  $, and $J^P=\frac{3}{2}^{-}(\frac{1}{2}^{+}\otimes1^{-})$.

Magnetic moments of the hadrons can be decomposed into two parts:
\begin{equation}
\mu = \mu_{spin}+ \mu_{orbital}, \label{TotMagMom}
\end{equation}
where  $\mu_{spin}$ is the spin magnetic moment and  $\mu_{orbital}$ is the orbital magnetic moment. We will consider S wave molecular picture in this work. Therefore there is no orbital magnetic moment $\mu_{orbital}$ between baryons and mesons. The operators of the corresponding magnetic moments in Eq. (\ref{TotMagMom}) can be written as
\begin{eqnarray}
\hat{\mu}_{spin} = {\sum_{i}}{\frac{q_{i}}{2m_{i}}}\hat{\sigma}_{i},
\end{eqnarray}
where $q_i$, $m_i $, and $\hat{\sigma}_{i}$ denote charge, mass, and Pauli’s spin matrix of the $i$th quark, respectively. Armed with these considerations, the total magnetic moment of a pentaquark state in molecular picture can be written as 
\begin{equation}
\hat{\mu} = \hat{\mu}_{B} + \hat{\mu}_{M},
\end{equation}
where the subscripts $B$ and $M$ represent baryon and meson, respectively.

\section{Results and Discussion}\label{numerical}
In the numerical analysis, we use three different sets of parameters for the constituent mass values which are listed in Table \ref{tab:input}. In this way, we can have an idea about how the constituent quark model parameters effect the results. Using these numerical values, we collect our results for the magnetic moments of $\Xi_b(6227)$ in Table \ref{tab:results}.

\begin{table}[h]
\caption{\label{tab:input}Constituent quark mass values. Mass values are given in unit of MeV.}
\begin{ruledtabular}
\begin{tabular}{cccc}
Quark & Set A \cite{Lichtenberg:1976fi} & Set B \cite{Dhir:2013nka} & Set C \cite{Karliner:2014gca}\\
\hline
$m_u=m_d$& 336 & 362  & 363 \\

$m_s$& 540  & 539& 538  \\

$m_b$& 4700  & 5043 & 5044 

\end{tabular}
\end{ruledtabular}
\end{table}

\begin{table}[h]
\caption{\label{tab:results}The magnetic moments of the  S wave $\Xi_b(6227)$ molecular pentaquark. The $J_{b} ^{P_{b}}\otimes J_{m} ^{P_{m}} $ represents corresponding angular momentum and parity of baryon and meson, respectively. The results are given in unit of nuclear magnetic moment $\mu_N$. }
\begin{ruledtabular}
\begin{tabular}{cccccc}
Multiplet &  $J_{b} ^{P_{b}}\otimes J_{m} ^{P_{m}} $& $I(J^P)$& Set A & Set B  & Set C \\
\hline\\
$8_{1f}$& $\frac{1}{2}^{+} \otimes 0^{-}$& $1{(\frac{1}{2})}^{-}$ & 0.72  & 0.66 & 0.66
\\ \\ \hline \\

& $\frac{1}{2}^{+} \otimes 1^{-}$& $1{(\frac{1}{2})}^{-}$ & -2.10 & -1.95& -1.94
\\ \\ \hline \\

& $\frac{1}{2}^{+} \otimes 1^{-}$& $1{(\frac{3}{2})}^{-}$ & -1.36 &  -1.27 & -1.27
\\ \\ \hline \hline \\

$8_{2f}$&$\frac{1}{2}^{+} \otimes 0^{-}$& $1{(\frac{1}{2})}^{-}$ &  -0.066  & -0.062 & -0.062 \\ \\ \hline \\

& $\frac{1}{2}^{+} \otimes 1^{-}$& $1{(\frac{1}{2})}^{-}$ & -0.76 & -0.77& -0.77
\\ \\ \hline \\

& $\frac{1}{2}^{+} \otimes 1^{-}$& $1{(\frac{3}{2})}^{-}$ & -2.50 &  -2.37 & -2.36

\end{tabular}
\end{ruledtabular}
\end{table}

As can be seen from the table, all the magnetic moment results are accessible in the experiments except $\frac{1}{2}^{+} \otimes 0^{-}$  of $8_{2f}$ flavor representation which can be challenging to measure. We observe that giving angular momentum to the meson in the molecular picture of $\Xi_b(6227)$ state changes magnetic moment significantly. For example using parameters of Set A, in the $8_{1f}$ flavor representation $\frac{1}{2}^{+} \otimes 0^{-}$ has a magnetic moment of $0.72 \mu_N$. Giving angular momentum to meson changes magnetic moment to  $-2.10 \mu_N$ of $\frac{1}{2}^{+} \otimes 1^{-}$. In the $8_{2f}$ flavor representation $\frac{1}{2}^{+} \otimes 0^{-}$ has a magnetic moment of $-0.066 \mu_N$, whereas $\frac{1}{2}^{+} \otimes 1^{-}$ has a magnetic moment of $-0.76 \mu_N$. This observation is also valid for parameters of Set B and Set C. Therefore the  magnetic moments of the same quantum numbers of related state can distinguish the inner structure of $\Xi_b(6227)$ state. Another observation is that, the constituent mass of $b$ quark does not heavily effect the magnetic moment results. This outcome can be seen in the results of $\frac{1}{2}^{+} \otimes 0^{-}$ and $\frac{1}{2}^{+} \otimes 1^{-}$ of the $8_{1f}$ and $8_{2f}$ representations. Our results suggest that value of the magnetic moment is determined by the light quarks.

In both $8_{1f}$ and $8_{2f}$ flavor representations, all the results are negative except $\frac{1}{2}^{+} \otimes 0^{-}$ in $8_{1f}$ which is positive. Whether positive of negative, the results of the magnetic moment is related to inner structure. Ref. \cite{Ozdem:2021vry} calculated magnetic moment of $\Xi_b(6227)$ in molecular picture as $\mu_{\Xi_b}=0.12 \pm 0.03 \mu_N$ using light-cone QCD sum rule formalism. Our results are not compatible with this work. The reason for this could be that the interpolating current used in related work is not a definite isospin eigenstate. In Ref. \cite{Ozdem:2024jty}, the magnetic moments of $\bar D \Sigma_c$, $\bar D^{*} \Sigma_c$, and $\bar D \Sigma_c^{*}$ pentaquarks are studied by the same author. It is shown that using definite isospin eigenstate interpolating current changed results significantly compared to earlier work. We expect that same conclusion could happen in $\Xi_b(6227)$ state.
 
In molecular scenario, it is possible to excite $\bar{K}$ meson compared to $\Sigma_b$ in $8_{1f}$ and $\Lambda_b$ in $8_{2f}$. This is due to the fact that the mass of the $\bar{K}$ meson is at the order of 490 MeV whereas masses of $\Sigma_b$ and $\Lambda_b$ baryons are at the order of 6000 MeV. If the $\Xi_b(6227)$ state has a P wave excitation, this could result in the shape of it which can be justified by observing magnetic moment experimentally.

\section{Final Remarks}\label{final}
Magnetic moments are important physical observables of hadrons and reflect their inner structures. Whatever being conventional or exotic hadrons, experimental observation of magnetic moment could shed light not only on the nature of the controversial and not yet fully understood exotic states but also conventional hadrons.  

In this present work, we study magnetic moments of $\Xi_b(6227)$ state in molecular picture using quark model. We use three constituent quark model parameters in order to observe how input parameters change magnetic moment results. The prevalent assumption of $\Xi_b(6227)$ state as exotic hadron is a molecular state of $\Sigma_b \bar{K}$. In addition to this assumption, we also study magnetic moments of $\Lambda_b \bar{K}$ state. 

We observe that giving angular momentum to meson in the $\Xi_b(6227)$ pentaquark molecular state change magnetic moment significantly. Furthermore we observe that the $b$ quark mass do not heavily change the results of magnetic moment. Aforementioned discussions indicate that the magnetic moment of the $\Xi_b(6227)$ state is determined by the light quarks.

Lattice QCD study of magnetic moment can present more information about the nature of $\Xi_b(6227)$. In Ref. \cite{Hall:2014uca} mass and magnetic moment of $\Lambda(1405)$ state have been calculated in molecular picture using lattice QCD. Light quark contributions to magnetic moment of $\Lambda(1405)$ in molecular picture was studied by lattice QCD \cite{Hall:2016kou}. Similar analysis can be conducted in lattice QCD for $\Xi_b(6227)$. 

To sum up, we obtained magnetic moments of $\Xi_b(6227)$  considering it as a molecular state. As a result of increasing luminosity, collaborations like PANDA, LHCb, BESIII, Belle II and so on may measure magnetic moment of $\Xi_b(6227)$. Experimental measurement will clarify the nature of $\Xi_b(6227)$. We hope that our study may be useful for experimental endeavours.

\bibliography{6227Baryon}

\end{document}